\documentclass[aps,superscriptaddress,showpacs,showkeys]{revtex4}
\usepackage{amssymb}
\begin{document}
\def\Barcelo{Barcel\'o}
\title{Analogue models for FRW cosmologies}
\author{Carlos \Barcelo}
\email[]{carlos.barcelo@port.ac.uk}
\homepage[http://www.tech.port.ac.uk/staffweb/barceloc/]{}
\affiliation{Institute of Cosmology and Gravitation,
University of Portsmouth, Portsmouth PO1 2EG, England}
\author{Stefano Liberati}
\email[]{liberati@physics.umd.edu}
\homepage[http://www2.physics.umd.edu/\~{}liberati]{}
\affiliation{Physics Department, University of Maryland, College Park,
MD 20742--4111, USA} 
\author{Matt Visser}
\email[]{Matt.Visser@mcs.vuw.ac.nz}
\homepage[http://www.mcs.vuw.ac.nz/\~{}visser]{}
\affiliation{School of Mathematical and Computing Sciences,
Victoria University of Wellington, New Zealand}
\date{\small 31 March 2003; \LaTeX-ed \today}
\begin{abstract}
  
  It is by now well known that various condensed matter systems may be
  used to mimic many of the kinematic aspects of general relativity,
  and in particular of curved-spacetime quantum field theory. In this
  essay we will take a look at what would be needed to mimic a
  cosmological spacetime --- to be precise a spatially flat FRW
  cosmology --- in one of these analogue models. In order to do this
  one needs to build and control suitable time dependent systems. We
  discuss here two quite different ways to achieve this goal.  One
  might rely on an explosion, physically mimicking the big bang by an
  outflow of whatever medium is being used to carry the excitations of
  the analogue model, but this idea appears to encounter dynamical
  problems in practice. More subtly, one can avoid the need for any
  actual physical motion (and avoid the dynamical problems) by instead
  adjusting the propagation speed of the excitations of the analogue
  model. We shall focus on this more promising route and discuss its
  practicality.

\end{abstract}

\maketitle
\def\g{\kappa}
\def\half{{1\over2}}
\def\L{{\mathcal L}}
\def\S{{\mathcal S}}
\def\d{{\mathrm{d}}}
\def\x{{\mathbf x}}
\def\v{{\mathbf v}}
\def\im{{\rm i}}
\def\etal{{\emph{et al\/}}}
\def\det{{\mathrm{det}}}
\def\tr{{\mathrm{tr}}}
\def\ie{{\emph{i.e.}}}
\def\bnabla{\mbox{\boldmath$\nabla$}}
\def\Box{\kern0.5pt{\lower0.1pt\vbox{\hrule height.5pt width 6.8pt
    \hbox{\vrule width.5pt height6pt \kern6pt \vrule width.3pt}
    \hrule height.3pt width 6.8pt} }\kern1.5pt}
\def\HRULE{{\bigskip\hrule\bigskip}}
\def\be{\begin{equation}}
\def\ee{\end{equation}}
\def\implies{\Rightarrow}

\center{\it This essay received an ``honorable mention''
  in the 2003 Essay Competition\\ of the Gravity Research Foundation.}
\widetext
\section{Motivation}
It is by now well known that various condensed matter systems may be
used to mimic many of the kinematic aspects of general relativity, and
in particular of curved-spacetime quantum field
theory~\cite{unruh,unexpected,analog,book}. Prior work has focussed
largely on black holes and event horizons, but there are also
interesting cosmological issues that can be addressed. In this essay
we will take a look at what would be needed to mimic a cosmological
spacetime --- to be precise a spatially flat FRW cosmology --- in one
of these analogue models. There are two quite different ways of trying
to achieve the same goal:
\begin{itemize}
\item One might rely on an explosion, physically
  mimicking the big bang by an outflow of whatever medium
  is being used to carry the excitations of the analogue
  model (see e.g.~\cite{FeFi}). Unfortunately, this idea
  appears to encounter dynamical problems in practice,
  problems not inherent to the type of geometry being
  reproduced.
  
\item More subtly, one can avoid the need for any actual
  physical motion (and avoid the dynamical problems) by
  instead adjusting the propagation speed of the
  excitations in the analogue model.
\end{itemize}
The physical metric we will be trying to emulate is that of a
spatially flat FRW cosmology
\be
\d s^2_{\mathrm{FRW}} = - c^2\; \d t^2 + a(t)^2 \; \d \vec x^2,
\label{E:frw}
\ee
where $a(t)$ is the scale factor of the universe as a function of
time.  In contrast the analogue models generically provide effective
metrics of the form~\cite{unruh,unexpected,analog,book}
\be
\d s^2_{\mathrm{effective}} =  {\rho(t,\vec x)\over c_s(t,\vec x) } \;
\left\{- [c_s^2(t,\vec x) -v^2(t,\vec x)] \; \d t^2 
- 2 \; \vec v(t,\vec x) \cdot dt \;\d\vec x 
+  \; \d \vec x^2 \right\}.
\label{E:effective}
\ee
Here $\vec v(t,\vec x)$ is the physical velocity of the medium, $
c_s(t,\vec x)$ is the propagation speed of whatever excitations we are
interested in studying, and the conformal prefactor $ {\rho(t,\vec x)/
  c_s(t,\vec x) }$ depends both on the dimensionality of spacetime and
[to some extent] on the specific choice of the analogue model. In the
form presented above the conformal factor is appropriate to either
ordinary sound in a classical
fluid~\cite{unruh,unexpected,analog,book}, or to phonons in a
BEC~\cite{bec}, both in (3+1) dimensions.

We now want to consider the fundamental question --- given the
availability of ``effective metrics'' of type (\ref{E:effective}), how
close can one get to reproducing a FRW metric of type (\ref{E:frw}).
That this question is non-vacuous can be deduced from the observation
that in the black hole context ``effective metrics'' of type
(\ref{E:effective}) never exactly reproduce the Schwarzschild
geometry~\cite{analog}, they can do so only up to an overall conformal
factor. We wish to check, in particular, if the same phenomenon shows
up in a cosmological context.

\section{Explosions}
As mentioned in the introduction, one plausible approach is to rely on
an actual physical explosion in the medium to mimic the big bang~\cite{FeFi}.
Start with the FRW metric $(\ref{E:frw})$ and substitute
\begin{equation}
\vec z = b(t) \; \vec x; 
\qquad 
\d \vec z = b(t) \; \d \vec x + \dot b(t) \; \vec x \; \d t; 
\qquad
 b(t) \; \d \vec x = \d \vec z -  {\dot b(t)\over b(t)} \;  \vec z \; \d t;
\end{equation}
Then, introducing a Hubble-like parameter,
\begin{equation}
H_b(t) =  {\dot b(t)\over b(t)}, 
\end{equation}
the FRW metric $(\ref{E:frw})$ is transformed to 
\begin{equation}
\d s^2_{\mathrm{FRW}} = 
- \left[c^2 - {a^2 \over b^2}\; H_b(t)^2 \; \vec z^2\right] \; \d t^2 
- 2 \; {a^2 \over b^2} \; H_b(t)\; \vec z \cdot \d \vec z \; \d t   
+  {a^2 \over b^2} \; \d \vec z^2.
\end{equation}
This metric is equivalent to the required ``effective metric'' form
[equation $(\ref{E:effective})$], provided we take
\begin{equation}
\vec v(t,\vec z) \leftrightsquigarrow H_b(t) \;\vec z;  
\qquad 
c_s  \leftrightsquigarrow {b \over a}\; c;
\qquad 
{\rho\over c_s} \leftrightsquigarrow {a^2 \over b^2}.
\qquad
\label{equivalences}
\end{equation}
The continuity equation for the medium, 
\begin{equation}
\dot\rho + \vec\nabla\cdot(\rho \; \vec v) = 0,
\end{equation}
particularized to the flow field $\vec v(t,\vec z) \leftrightsquigarrow
H_b(t) \;\vec z$ implies
\begin{equation}
\dot\rho + 3 \rho\; H_b(t) = 0; 
\qquad \implies \qquad 
\rho(t)\propto {1\over b^3(t)}.
\label{eq:rho}
\end{equation}
Using the last equivalence in Eq.~(\ref{equivalences}) we can now fix
the behaviour of $c_s(t)$ as a function of $a(t)$ and $b(t)$
\be
c_s(t)\propto \frac{1}{a(t)^2 b(t)}.
\label{eq:cs}
\ee
Equations~(\ref{eq:rho}) and (\ref{eq:cs}),
together with the first equivalence in
Eq.~(\ref{equivalences}), completely fix the relation
between the hydrodynamical quantities $(\rho(t),\vec
v(t,\vec{z}), c_s(t))$ and the cosmological solution
parameters $(a(t),b(t),c)$.

Although the ``explosion route'' just discussed seems promising, one
should note that because of the linearly rising velocity field $\vec v
= H_b \vec z$, this particular realization of a FRW effective geometry
is guaranteed to possess an apparent horizon, a spherical surface in
which the speed of the fluid surpasses the speed of sound.  From a
dynamical point of view, this might introduce many practical problems
not intrinsically inherent to the type of geometries we are trying to
reproduce.  In this sense, we view the use of an exploding medium as
not being a particularly useful analogue for an expanding FRW
universe.

\section{Varying propagation speed}

A much better analogue of an expanding FRW universe can be obtained by
keeping the medium at rest and instead varying the propagation speed
of the excitations. In this case the continuity equation implies that
$\rho(t,\vec x)$ is a constant, so that we can rescale the metric by a
constant factor $c_0/\rho$ (where $c_0$ is any convenient reference
speed) obtaining
\be
\d s^2_{\mathrm{effective}} 
=  {c_0 \over c_s(t) } \;
\left\{- c_s^2(t) \; \d t^2 
+   \d \vec x^2 \right\} 
=  -{ c_0 \; c_s(t)\; \d t^2} +  { c_0 \over c_s(t)}\; \d \vec x^2.
\label{E:effective2}
\ee
Now introduce a pseudo-time $\tau$, related to laboratory time via
$\d\tau = \d t \; \sqrt{c_s(t)/c_0}$.  Then
\be
\d s^2_{\mathrm{effective}} =  
-c_0^2\; {\d\tau^2} + {c_0 \over c_s(t)}\; \d \vec x^2.
\label{E:effective3}
\ee
Consequently we completely reproduce a FRW cosmology provided we identify
\begin{equation}
c_0 \; \d\tau \leftrightsquigarrow c \; \d t_{\mathrm{FRW}}; 
\qquad  
{c_0\over c_s(t)}\leftrightsquigarrow a(\tau)^2.
\end{equation}
That is, an expanding universe corresponds in the effective geometry
to a decreasing propagation speed. A tricky point is that $c_s$ is
still presented in terms of laboratory time $t$ as the speed $||\d\vec
x/\d t||$. However in terms of the pseudo-time $\tau$ the excitations
propagate at speed
\begin{equation}
\bar c_s(\tau) 
= \left|\left|{\d\vec x\over\d \tau}\right|\right| 
= {\d t\over\d \tau} \;  \left|\left|{\d\vec x\over\d t}\right|\right| 
= \sqrt{c_0\over c_s(t)} \; \; c_s(t) 
=  \sqrt{c_0 \; c_s(t)}.
\end{equation}
Having obtained the general analogue of the flat FRW metric it is
interesting to investigate the analogue equivalent of the inflationary
solution.  We can start by re-writing the analogue Hubble factor as
\be 
H
=
{{a^{\prime}(\tau)}\over {a(\tau)}} 
= 
{1\over2} \;\sqrt{c_s\over c_0}\; \dot{a} \; {\d t\over\d \tau} 
= 
-{1\over2}\; \sqrt{{c_0\over c_s}}\;{\dot{c_s}\over c_s}.  
\ee 
where the prime represents derivative with respect to the pseudo-time
and the dot with respect to the laboratory time. The inflationary
solution is then easily obtained by using a exponential law in
pseudo-time $\tau$
\be
a(\tau)=  e^{H\tau}; 
\qquad
a(t)={H\; t}; 
\ee
corresponding to a power law in physical time for the speed of sound
(measured in physical time)
\be
c_s(t)= \frac{c_0}{a^2(\tau)}=\frac{c_0}{H^2\,t^2}.
\ee
In summary the nice feature of the variable propagation speed route to
a FRW analogue is that it is relatively clean; there is no moving
medium which might impact the physical boundaries of any experimental
apparatus and one can instead focus on what we feel is the central
issue --- how to physically manipulate the propagation speed $c_s$ (or
its pseudo-time equivalent $\bar c_s$).

\section{Suitable physical mechanism: Feshbach resonance in a BEC}

{From} the preceding discussion it is clear that one attractive route
to simulating a FRW cosmology in an effective geometry is by rapid
manipulation of the propagation speed of whatever excitations we might
wish to focus attention on. But how is that to be accomplished? There
is one particular medium, currently the center of much experimental
interest, for which an appropriate mechanism has been demonstrated to
exist. Here we are referring tho the use of Feshbach
resonances~\cite{feschbach} in Bose--Einstein condensates [BECs].

BECs are promising analogue systems of gravity. They have been
extensively studied~\cite{book,Garay1,bec,laval} in relation to the
possibility of simulating black hole geometries and Hawking radiation.
The basic equation used in describing the condensate is the
Gross-Pitaeveskii [GP] equation
\be
\label{E:gp}
 - \im \hbar \; \frac{\partial }{\partial t} \psi(t,\x)= \left (
 - {\hbar^2\over2m}\nabla^2 
 + V_{\rm ext}(\x)
 + \kappa \; \left| \psi(t,\x) \right|^2 \right) \psi(t,\x);
\qquad 
\kappa={4\pi \sigma \hbar^2\over m}
\ee
Here $\psi(x,t)$ is the (classical) wave function of the condensate,
($|\psi(x,t)|^2=n=$the particle density of the condensate), $V_{\rm
  ext}$ is the trapping potential and $\sigma$ is the $s$-wave
scattering length for the atoms (which have mass $m$). [Actually it is
conventional to use the symbol $a$ for the scattering length of a BEC
condensate. We adopt this unusual notation to avoid any confusion with
the scale factor of the FRW metric.]  The generic analogue model can
be easily be obtained by considering the propagation of excitations in
the condensate.  These are described by a wave function which closely
resembles that of a scalar field in a curved spacetime described by
the metric (\ref{E:effective}) with $\rho$ replaced by $n\;m$ and
where $c_s^2=\kappa n/m$ (see~\cite{bec} for a detailed derivation).
The key point is that the propagation speed is proportional to the
scattering length
\be
c_s^2 \propto \sigma.
\ee

Let us now qualitatively explain we can simulate cosmological
expansion within this model. We start with a condensate in a
stationary state described by a constant background density in a
sufficiently large volume. This is a solution of the Gross-Pitaevskii
equation.  For this, one needs to have a potential that reproduces a
sufficiently large hard-walled box.  Using a Feshbach resonance it is
now possible to change at will the scattering length $\sigma$ in the
condensate. If one now decreases the value of the scattering length in
a sufficiently slow manner (the timescale will be discussed below),
then at the same time the value of the speed of sound will decrease.
The analysis of the previous Section then shows that this leads to an
effective FRW geometry. Fluctuations of the condensate will propagate
in an effective metric which is an analogue to a spatially flat FRW
geometry.

A subtle point is that when the time dependence is introduced in the
system by a time-varying scattering length one must be sure to work in
a regime where the background configuration is ``instantaneously''
reacting to theses changes. Thus we must assure that the GP equation
(from which the analogue gravity framework is derived) holds at each
instant of time.  The validity of the GP in describing the
Bose-Einstein condensate is related to the validly of several crucial
assumptions, generally stated to be the ``mean-field'' approximation
and the dilute gas approximation.  It is nevertheless important to
note that in a dynamical situation a third approximation, which we can
call ``Markovian'' approximation, is implicitly assumed.

The Markovian approximation is related to the fact that in dynamical
situations the two-body time-dependent scattering matrix can have a
complicated form due to the ``memory'' of the system. In these
situations the system is described by a GP-like equation where the
interaction term includes a ``delay term'' described by an integration
over time. The necessary assumption in order to have a Markovian
description of the dynamics (which together with the two previous
approximations leads to the GP equation) is then that the timescales
on which external parameters are changing are longer that the two-body
collisional duration. That is, longer than the timescale over which a
single interaction happens.  (Reduced to the bare bones, we are asking
that the scattering length does not change significantly during the
period when a pair of atoms are interacting.)

We can estimate the two-body collision time by a simple calculation.
All we need is the typical size of the region of strong interaction of
two atoms in the condensate, and the typical speed with which they
move. The first quantity can be assumed to be of the order of the Van
der Waals length.  This length is basically the size of the region of
strong interaction: For $r \ll \lambda_{vdW}$ the scattering wave
function oscillates rapidly due to the strong interaction potential.
In alkali ground state interactions, $\lambda_{vdW}$ is typically of
order of few nanometers, $\lambda_{vdW}\approx 1$~nm.

Regarding the typical speed of the atoms, this is set by the
de~Broglie momentum generated by the trap confinement: $p=h/R$ and
$\bar{v}=p/m$. We shall assume a trap of typical size of ten microns.
For a typical Bose-Einstein condensate one then gets an interaction
timescale
\be
t_{i}=\frac{\lambda_{vdW}}{\bar{v}}=\frac{\lambda_{vdW}\;m\;R}{h}
\approx 10^{-6} \mbox{ s},
\label{eq:tin}
\ee
so it would appear that a microsecond is the shortest timescale
allowed for the change in $\sigma$.

\section{Analogue cosmological particle creation}

Now that the above analysis has shown that there is a regime for which
a varying scattering length can be used to simulate a FRW-like
effective metric one may wonder about the behaviour of the
fluctuations on such a time-dependent background.

The equation satisfied by the quantum fluctuations is, in the acoustic
approximation (that is for long wavelengths), that of a massless
scalar field over an expanding background and therefore, it will lead
to cosmological pair production of particles. Interestingly it was
recently demonstrated~\cite{Un-Sc} that not all analogue models of
gravity are suitable for simulating particle creation from the quantum
vacuum. Indeed, it may happen that even if the classical equation for
the perturbations on the background resembles that for a field in
curved spacetime, nonetheless one might fail to mimic quantum particle
creation because the commutation relations of the analogue field are
not the correct one for the physical quantum field.  In this regard it
is useful to note that it can be explicitly shown~\cite{Un-Sc} that in
the case of BECs the structure of the commutator is the correct one.

We then expect to be able to produce quantum excitations with several
frequencies, typically determined by the rapidity of the change in the
scattering length. In particular if $t_{\mathrm{min}}$ is the shortest
timescale over which we can physically drive the system then
$\nu_{\mathrm{peak}} \approx 1/t_{\mathrm{min}}$ is the peak frequency
of the created quasi-particles spectrum. It is also important to check
that this frequency corresponds to a wavelength shorter than the
physical size of the condensate $R$. An easy way to do this is to
confront the interaction time $t_{i}$ given in Eq.~(\ref{eq:tin}) with
the crossing time of the condensate, $t_{\rm size}=R/c$.
\begin{equation}
\frac{t_{i}}{t_{size}}=
\frac{\lambda_{vdW}\; m c}{h}= 
\frac{2\pi \lambda_{vdW}}{\xi},
\end{equation}
where we have introduced the healing length of the condensate, defined
to be $\xi=\hbar/(m c_s)$.  For typical BEC system $\xi\approx
0.1$--$10\mu$m (assuming an average value of the scattering length of
a few nanometers) so
\begin{equation}
\frac{t_{i}}{t_{\rm size}}\approx 10^{-2}\mbox{--}10^{-4}.
\end{equation}
Thus there is a viable window of timescales (for the time
dependence of the scattering length) for which both the GP equation
holds, and the quasi-particles produced have wavelengths shorter
than the physical size of the condensate. 

The possibility of simulating cosmological particle creation is a
significant advance that deserves further investigation, and might be
remarkably important in the future. The simulation of inflationary
scenarios, and the relevant particle creation, could lead to a better
understanding of the generation of primordial inflationary
perturbations and their role in the generation of large scale
structure. Moreover it should be noted that the quasi-particles
generated in this way will be characterized by a phononic dispersion
relation $\omega^2=c_s^2\,k^2$ only at long wavelengths with respect
to the healing length of the condensate. In general for wavelengths
comparable with the healing length of the condensate the Bogoliubov
dispersion relation
\begin{equation}
\omega=\sqrt{c_s^2 k^2+\left({\hbar \over 2 m}\;k^2\right)^2},
\label{E:bogo}
\end{equation}
will hold~\cite{broken}. This is particularly interesting due to the
recent intense debate about the detectability of inflationary spectrum
due to these kind of dispersion relations~\cite{Bran-Mart,Pare}.

\section{Conclusions and Prospects}

Our major conclusions are four-fold:
\begin{itemize}
\item At a theoretical level, it is clear that any mechanism for
  changing the propagation speed in a stationary medium is from a
  mathematical perspective equivalent to working in an expanding (or
  collapsing) spatially flat FRW universe.
  
\item At an experimental level, the use of a Feshbach resonance in a
  Bose--Einstein condensate yields a way of influencing the scattering
  length, and hence the propagation speed, without changes in
  condensate density --- and more importantly without violating the
  approximations made in deriving the Gross-Pitaevskii equation on
  which the entire ``effective metric'' approach to phonon propagation
  in BECs is based.
  
\item At a very practical engineering level, the relevant parameters
  seem to be well within our technological horizon.
  
\item Building an analogue FRW cosmology (suitable for testing
  semiclassical quantum effects) seems considerably less problematic
  than building an analogue black hole~\cite{book,Garay1,laval}.
\end{itemize}
In summary: The prospects for direct laboratory simulation of an
expanding universe, and consequent cosmological particle production
are very good.  Given the relatively small number of experimental
tests of curved space quantum field theory, any progress along these
lines is important. Our long range goal is to turn at least some
aspects of cosmology into a laboratory science, not just an
observational science.


\end{document}